\documentclass{article}

\usepackage{arxiv}

\usepackage[utf8]{inputenc} 
\usepackage[T1]{fontenc}    
\usepackage{authblk}
\usepackage{cite}
\usepackage{float}
\usepackage{comment}
\usepackage{hyperref}       
\usepackage{url}            
\PassOptionsToPackage{hyphens}{url}\usepackage{hyperref}
\usepackage{booktabs}       
\usepackage{amsfonts}       
\usepackage{nicefrac}       
\usepackage{microtype}      
\usepackage{lipsum}
\usepackage{longtable}
\usepackage{subfig}
\usepackage{graphicx}
\usepackage{doi}
\graphicspath{ {./images/} }
\title{Contributions of the Petabyte Scale Sequence Search Codeathon toward efforts to scale sequence-based searches on SRA
}

\date{}

\author[1,*]{Priyanka Ghosh}
\author[2]{Kjiersten Fagnan}
\author[1]{Ryan Connor}
\author[1]{Ravinder Pannu}
\author[3,18]{Travis J. Wheeler}
\author[5]{Mihai Pop}
\author[4]{C. Titus Brown}
\author[4]{Tessa Pierce-Ward}
\author[5,10]{Rob Patro}
\author[5,6]{Jacquelyn S. Michaelis}
\author[1]{Thomas L. Madden}
\author[1]{Christiam Camacho}
\author[8]{Olaitan I. Awe}
\author[9]{Arianna I. Krinos}
\author[11]{René KM Xavier}
\author[12]{Rodrigo Ortega Polo}
\author[3,18]{Jack W. Roddy}
\author[1,26]{Adelaide Rhodes}
\author[7,25]{Alexander Sweeten}
\author[23]{Adrian Viehweger}
\author[14]{Barış Ekim}
\author[5]{Harihara Subrahmaniam Muralidharan}
\author[15]{Amatur Rahman}
\author[16]{Vin\'{i}cius W. Salazar}
\author[17]{Andrew Tritt}
\author[3,18]{Thomas Colligan}
\author[19]{Katrina Kalantar}
\author[3,18]{Genevieve R. Krause}
\author[20]{Taylor Reiter}
\author[18]{George Lesica}
\author[21]{Artem Babaian}
\author[13]{Victor Lin}
\author[22]{Sergey Madaminov}
\author[1]{Vadim Zalunin}
\author[1]{David M. Kristensen}
\author[1]{Alexa Salsbury}
\author[1,24]{Daniel P. Rice}
\author[1]{J. Rodney Brister}

\affil[1]{National Center for Biotechnology Information, National Library of Medicine, National Institutes of Health,
Bethesda MD, USA;}
\affil[2]{Joint Genome Institute, Lawrence Berkeley National Laboratory, Berkeley CA, USA }
\affil[3]{Department of Pharmacy Practice and Science, R. Ken Coit College of Pharmacy, University of Arizona, Tucson AZ, USA}
\affil[4]{University of California Davis, Davis CA, USA}
\affil[5]{Department of Computer Science, University of Maryland at College Park, College
Park, MD, USA}
\affil[6]{Center for Bioinformatics and Computational Biology, University of Maryland, College Park, MD USA}
\affil[7]{Center for Genomics and Data Science Research, National Human Genome Research Institute, National Institutes of Health, Bethesda MD, USA}
\affil[8]{African Society for Bioinformatics and Computational Biology, Cape Town, South Africa}
\affil[9]{Department of Earth, Environmental, and Planetary Sciences, Brown University, Providence, RI USA}
\affil[10]{Ocean Genomics Inc., Pittsburgh, PA USA}
\affil[11]{Florida Atlantic University Harbor Branch Oceanographic Institute, Fort Pierce Florida, USA}
\affil[12]{Lethbridge Research and Development Center, Agriculture and Agri-Food Canada, Lethbridge, Canada}
\affil[13]{Unaffiliated}
\affil[14]{Computer Science and Artificial Intelligence Laboratory, Massachusetts Institute of Technology, Cambridge, MA USA}
\affil[15]{Department of Computer Science and Engineering, The Pennsylvania State University, University Park, PA USA}
\affil[16]{Melbourne Bioinformatics, University of Melbourne, Parkville, VIC, Australia}
\affil[17]{Applied Mathematics and Computational Research Division, Lawrence Berkeley National Laboratory, CA USA}
\affil[18]{Department of Computer Science, University of Montana, Missoula, MT USA}
\affil[19]{Chan Zuckerberg Initiative, Redwood City, CA USA}
\affil[20]{Department of Population Health and Reproduction, School of Veterinary Medicine, University of California Davis, CA USA}
\affil[21]{Department of Molecular Genetics, University of Toronto, Ontario, Canada}
\affil[22]{Department of Computer Science, Stony Brook University, NY USA}
\affil[23]{Institute of Medical
Microbiology and Virology, University Hospital Leipzig, Leipzig Germany}
\affil[24]{SecureBio, Cambridge, MA USA}
\affil[25]{Department of Computer Science, Johns Hopkins University, Baltimore MD USA}
\affil[26]{Genedata, Inc., Lexington, MA, USA}
\affil[*]{Corresponding Author: priyanka.ghosh@nih.gov}

\begin{document}
\maketitle

\begin{abstract}
\newpage
The volume of biological data being generated by the scientific community is growing exponentially, reflecting technological advances and research activities. The National Institutes of Health’s (NIH) Sequence Read Archive (SRA), which is maintained by the National Center for Biotechnology Information (NCBI) at the National Library of Medicine (NLM), is a rapidly growing public database that researchers use to drive scientific discovery across all domains of life. This increase in available data has great promise for pushing scientific discovery but also introduces new challenges that scientific communities need to address. 

\par As genomic datasets have grown in scale and diversity, a parade of new methods and associated software have been developed to address the challenges posed by this growth. These methodological advances are vital for maximally leveraging the power of next-generation sequencing (NGS) technologies, but it can be difficult to make sense of their performance trade-offs (especially speed and accuracy). With the goal of laying a foundation for evaluation of methods for petabyte-scale sequence search,
the Department of Energy (DOE) Office of Biological and Environmental Research (BER), the NIH Office of Data Science Strategy (ODSS), and NCBI held a virtual codeathon `Petabyte Scale Sequence Search: Metagenomics Benchmarking Codeathon' on September 27 - Oct 1 2021, to evaluate emerging solutions in petabyte scale sequence search. The codeathon attracted experts from national laboratories including the Lawrence Berkeley National Laboratory (LBNL), research institutions including the Joint Genome Institute (JGI), and universities across the world to (a) develop benchmarking approaches to address challenges in conducting large-scale analyses of metagenomic data (which comprises approximately 20\% of SRA), (b) identify potential applications (i.e. use-cases) that benefit from SRA-wide searches and the tools required to execute the search, and (c) produce community resources made accessible to researchers with an interest in using the data i.e. a public facing repository with information to rebuild and reproduce the problems addressed by each team challenge. 

\end{abstract}


\section{Introduction}
\label{sec:intro}

The revolution in NGS technologies has provided profound insights into the genetic composition and variation within biological samples and has yielded a number of scientific breakthroughs including the sequencing of the human genome ~\cite{venter2001sequence,nurk2022complete}. This technological advancement 
has been supported by investments in fundamental research, technology, and infrastructure; the resultant decrease in nucleotide sequencing and data storage costs have led to an exponential increase in the volume of sequence data generated by the scientific community. Rapid evolution of NGS and associated methodologies presents significant challenges in management and analysis of large datasets and for extracting biologically or clinically relevant information. In this manuscript we discuss the contributions and efforts of the Petabyte Scale Sequence Search (PSSS) consortium to address the growing need for developing scalable sequence-based search algorithms/tools to enable scanning SRA\cite{shumway2010archiving} (akin to a BLAST\textsuperscript{\textregistered} search across the entire SRA) at petabyte scale, with the goal to tackle problems in functional genomics associated with metagenomic data with high-impact health outcomes and the development of computational workflows that support data discovery and reusability.

\subsection{SRA has reached the petabyte scale}
\label{sec:SraGrowth}

SRA is a diverse collection of NGS and 3rd generation sequencing data managed by the  NLM\cite{kodama2012sequence, katz2022sequence}. SRA provides a repository where data creators can share their raw sequence data with the scientific community. It holds both public data available to all researchers and controlled access data derived from human research studies for use by qualified biomedical investigators who agree in advance to use the data appropriately. Sequence data in the archive is linked to associated metadata that provides information about the sequenced sample and can be used to associate genetic information in SRA with phenotypic, clinical, and environmental attributes \cite{katz2021stat}. The archive is also part of the International Nucleotide Sequence Database Collaboration (INSDC), and SRA sequence data and associated metadata as well as analogous data submitted to the European Nucleotide Archive (ENA) and DNA Data Bank of Japan (DDBJ) are exchanged among member databases.

\begin{figure}[t]
        \centering
        \includegraphics[width=0.6\textwidth]{
        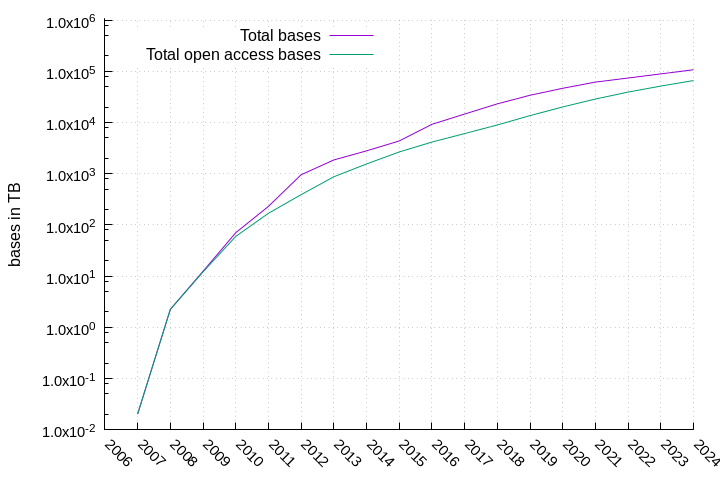
        }
        \caption{Growth of SRA. Databases like SRA are growing rapidly and are used extensively by scientific communities. As these databases grow, so do their potential scientific value, along with the complexity of ensuring ease of access.}
        \label{fig:sragrowth}
\end{figure}

\begin{figure}[tp]%
    \centering
    \subfloat[\centering Growth in the number of SRA runs and biosamples by release date. \label{fig:2a}]
    {{\includegraphics[width=8.3cm]{
    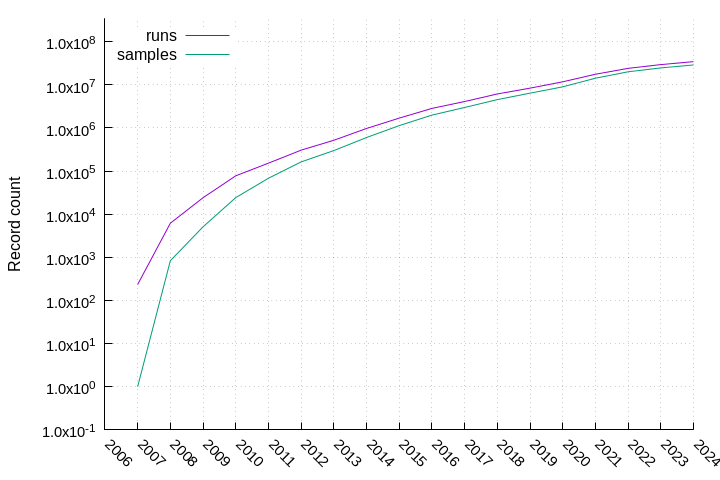
    } 
    }}%
    \qquad
    \subfloat[\centering Growth in the number of metagenomic runs and samples in SRA by release date. \label{fig:2b}]
    {{\includegraphics[width=8.3cm]{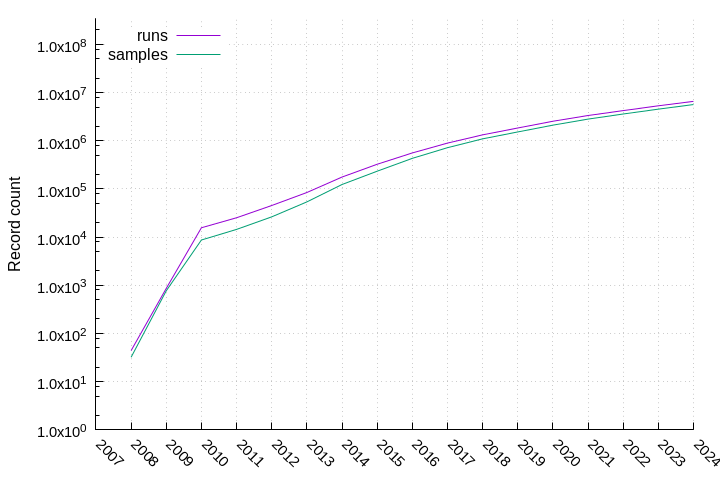
    }
    }}%
    \qquad
    \subfloat[\centering Individual SRA record size by year. \label{fig:2c}]
    {{\includegraphics[width=8.3cm]{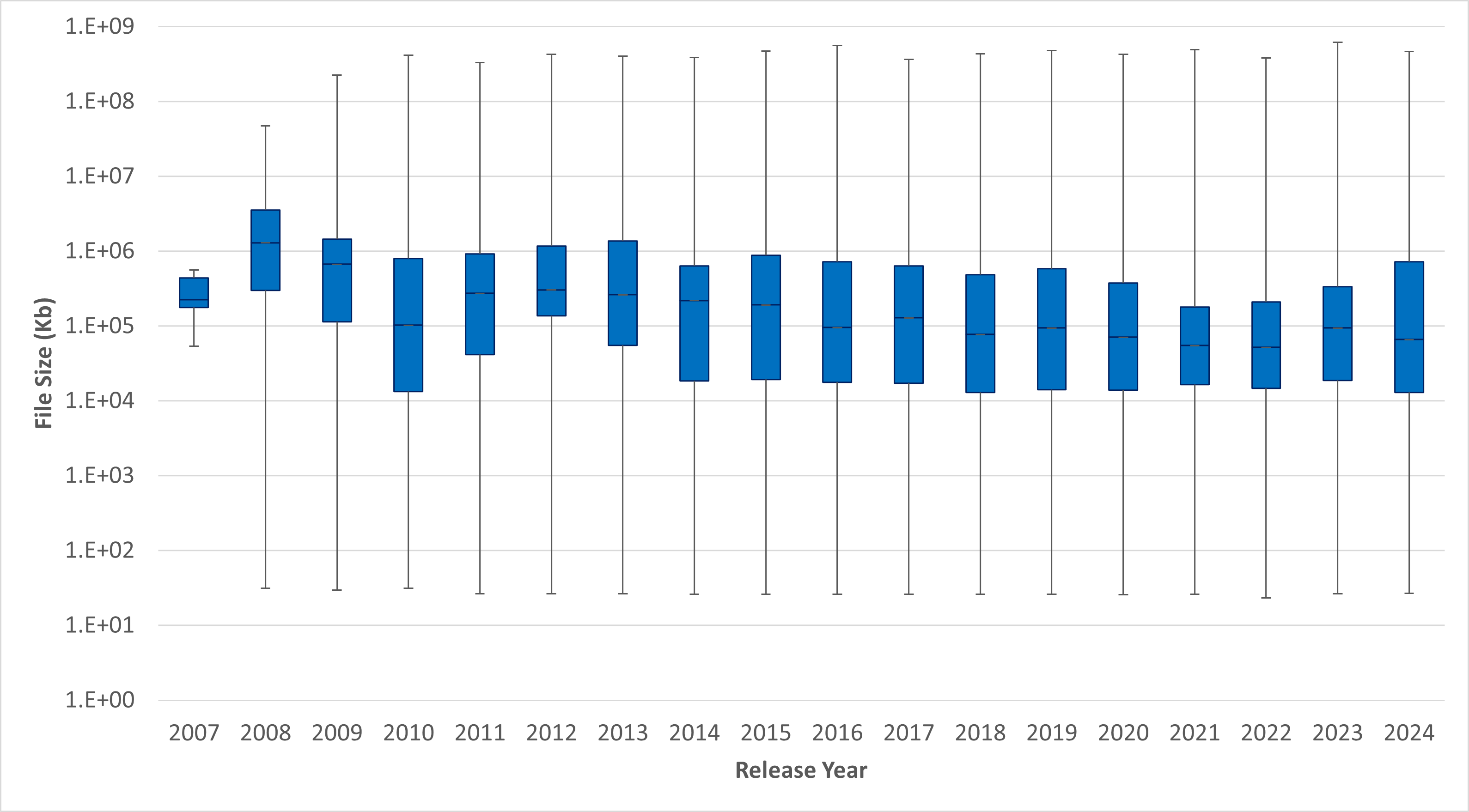}}}%
    \caption{SRA cumulative records growth over time. `runs' refer to SRA accessions and `samples' correspond to biosamples, a record in SRA that contain descriptions of biological source material used in experimental assays.}%
    \label{fig:SRArecordgrowth}%
\end{figure}


The growth in nucleotide sequencing data within the scientific community has led to an explosion in the size of SRA (as seen in Fig.~ \ref{fig:sragrowth}). 
There are now more than a total of 34.5 million submitted samples from all domains of life that together comprise more than 107 petabases corresponding to more than 38 petabytes of raw sequencing data in the repository 
as of December 2024. The number of publicly available samples is greater than 33.1 million and comprises of 66.9 petabases ($\sim$ 27 petabytes) of sequencing data. The growth is observed not only in the number of records (as seen in Fig. \ref{fig:2a}), but also in 
variation of record size (Fig. \ref{fig:2c}), diversity of organisms (Fig. \ref{fig:metagenomicsradata}), 
wider variety of data types, and in sequencing technologies used to generate the data. Given its vast size and diversity, SRA represents a vital resource for the scientific community, and researchers have developed bioinformatic tools and methods that support deep analysis of SRA data – everything from assembly and annotation of genomes, characterization of human pathogens, discovery of novel organisms and viruses, and association of genetic signatures from complex microbiome and metagenomic samples with environmental attributes, to prediction of the functional significance of rare human genetic variation.

To sustain future growth and facilitate the expanded use of the petabyte-scale SRA dataset, NIH partnered with Google Cloud Platform (GCP) \cite{GCP} and Amazon Web Services (AWS) \cite{AWS} in 2019, through the NIH Science and Technology Research Infrastructure for Discovery, Experimentation, and Sustainability (STRIDES) Initiative \cite{stridesini}. This partnership was part of a larger NIH STRIDES effort to build a cloud-enabled biomedical dataverse, with new processes, tools, and architecture to drive development of an equitable ecosystem that makes NIH-funded data findable, accessible, interoperable, and reusable (FAIR) \cite{wilkinson2016fair}. The entirety of SRA sequence dataset is now replicated on Google and Amazon cloud platforms, and associated BioProject~\cite{barrett2012bioproject, pruitt2011bioproject}, BioSample, and SRA metadata are available through Google Big Query and Amazon Athena services. The public SRA dataset is available in normalized format through the AWS Open Data Program, and data can be egressed without charge into cloud and local environments~\cite{sracosts}. 

Hosting SRA data in the cloud enables the techniques for analysis  at a comprehensive scale. Direct access to SRA data as normalized objects will enable more scalable and cloud-native tooling for processing and analyzing genomics datasets. Rather than copying terabytes or even petabytes of data into their own environments, researchers can perform analysis using cloud resources proximal to the data, with the added benefit of making workflows more reproducible and amenable to global research collaborations.

\subsection{Petabyte-scale sequence search will improve the scientific impact of SRA}
\label{sec:psss}

While staging on cloud platforms provides enhanced access to SRA, successful use of the archive demands the ability to search across the petabyte-scale dataset to locate samples of interest. Search in this context can be thought of as two discrete functions: (1) text-based search, where queries to are matched to metadata descriptors associated with deposited sequence samples, and (2) sequence-based search, where sequence strings are matched to the sequence content of a sample. Attribute-based search depends on accurate sample metadata and is hindered by missing, incorrect, or inconsistently labeled or formatted metadata. In contrast, sequence-based search is essentially agnostic to the sample information provided by data submitters and depends instead on the content of the sequence data itself.

Sequence-based search has been critical to the application of GenBank\textsuperscript{\textregistered}\cite{benson2013genbank} as a public genetic sequence repository. The Basic Local Alignment Search Tool (BLAST)~\cite{boratyn2013blast} provided a way for researchers to compare experimentally derived sequences to those in GenBank and to identify sequences of interest. BLAST supports sequence-based searches and can execute nucleotide-nucleotide (blastn), protein-protein (blastp) and protein-nucleotide (tblastn/blastx) searches. These different search modalities support the identification of sequences that have a broad range of sequence similarity to the query, from close homologues to distantly-related sequences.
Similar search functionality will be critical to the usability of SRA. On one hand, the immense scale of the database ($\sim$4,200 times the size of GenBank in bases, by the end of 2024) imposes unique challenges to sequence search tools. On the other hand, the same scale will provide unique scientific rewards if such tools are successfully implemented.

\subsubsection{The value of sequence-based search}
\label{sec:SeqSearch}

Searching an entire database like SRA enables researches to explore multiple biological use-cases that require finding samples for identifying new relationships yet unknown, such as tracking and surveillance of pathogen spread in public data, searches for small viruses, detecting sequences belonging to novel taxonomic classes, cataloging the diversity and prevalence of pathogenic species and virulence genes and aiding design of better detection assays, to name a few. 
Searches across SRA corpus would enable us to ask and answer a variety of new and interesting questions. For example,
recent work by Serratus \cite{edgar2022petabase} to recover 880,000 RNA-dependent RNA polymerase-containing sequences demonstrates the potential of SRA-wide sequence search to identify novel viruses. More recently, Logan~\cite{logan} utilized genome assembly over each publicly available SRA accession (as of December 2023) to produce the largest accessible repository of assembled sequencing data (unitigs and contigs) by using massively parallel cloud computing resources. However, no extant tool supports a one size fits all approach to sequence-based search across all of SRA, and search strategies vary with the specific use cases and underlying data. 


\subsubsection{Metadata Opportunities and Challenges }
\label{sec:MetadataChallenges}

When the scientist’s goal is discovery of what relevant data might be available in the repository, a direct search on sequence data circumvents issues with missing or inaccurate metadata. Improving the accuracy of metadata is challenging and is an active area of research, considering the possibility that metadata search could serve as an alternative to sequence similarity. 
Tools such as Metaseek ~\cite{hoarfrost2019sequencing} aggregate SRA metadata and preliminary investigation shows that many SRA metadata entries are incomplete. We are interested in the types of discovery that are possible if scientists can quickly determine all the datasets that contain a particular gene or genome. The metadata associated with those datasets may be inaccurate and that will decrease the utility in reusing that data. Current advances in the area of large language models may yield improvements in language alignment that make metadata-based search strategy more powerful by improving current and future metadata quality. Exploring the impact of metadata is outside the scope of this discussion or the benchmarks developed as part of the  codeathon.

\begin{figure}[t]
        \centering
        \includegraphics[width=0.7\textwidth]{
        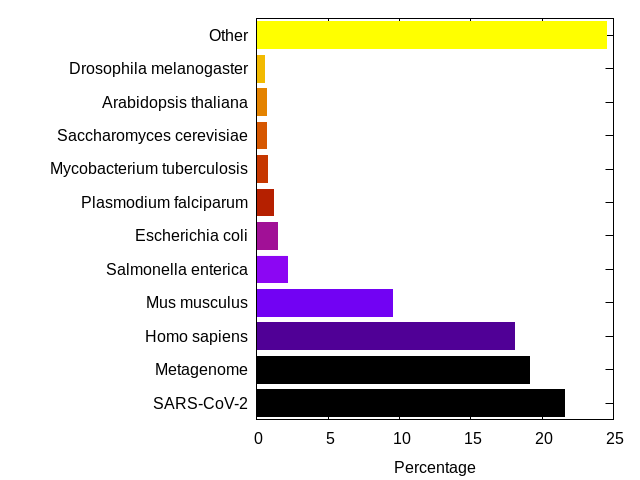
        }
        \caption{Percent SRA accessions by organism as of December 2024.}
        \label{fig:metagenomicsradata}
\end{figure}

\subsection{Goals of the Codeathon}
\label{codeathongoals}

The codeathon was focused on a particular scientific domain, metagenomics, wherein the goal lies in defining within the broad umbrella of sequence search - (1) a set of specific sequence search problems, (2) a set of performance benchmarks that can specify the parameters for success in solving each of the problems (i.e. measures of accuracy and computational performance), and finally (3) pipelines deployed on the cloud that enable benchmarking of existing tools and comparisons to newly developed tools, providing a common ground for understanding what approaches are potentially successful at solving these problems at large scale. We discuss the chosen challenge problems identified by each of the teams and their associated use-cases in greater detail in section \ref{sec:psss-codeathon}.

\subsection{Metagenomics}
\label{sec:MetagenomicsData}

A metagenome is a dataset that represents a collection of genomes co-existing in a sample, typically comprised of complex communities of viruses, bacteria, fungi, and other micro-eukaryotes (referred to as microbes) that inhabit most of the planet. The sample may be from the human gut, a marine estuary, or your office. The study of the microbial communities is important because many phenomena are driven not just by the actions of a single microbial species, but many working together. Metagenomes are large, complex, and challenging to analyze; they may contain the genomes of thousands of organisms, and the largest dataset sizes are on the order of 10TB. An increase in the number of metagenomic studies has resulted from a reduction in sequencing costs as well as more sophisticated data processing and analysis techniques. The number and size of metagenomic datasets being deposited and utilized in SRA is one of the fastest growing among other datasets. Further reasoning for focusing the codeathon on metagenomic data include:

\begin{itemize}
    \item A significant percentage ($\sim$20\%) of SRA data is comprised of metagenomic and metatranscriptomic runs as seen in Figs.~\ref{fig:2b}, \ref{fig:metagenomicsradata}. Enabling searches across the vast corpus of metagenomic content in SRA is key, and requires substantial computational resources to efficiently analyze datasets.
    \item Metagenomic datasets provide a unique set of challenges that include both known and unknown organisms within a sample that can contain upwards of thousands of different genomes with varying degrees of similarity, sometimes sharing genetic material, and occurring at vastly different abundances. In addition, data is often generated using different sequencing platforms with varying read lengths and quality leading to biases in detected communities. 
    \item The ability to identify organismal content and/or correlate it with sample attributes is critical to the use of metagenomic sequence datasets. Furthermore, many microbial species in metagenomic datasets are either novel or rare, thereby making accurate taxonomic classification difficult.

\end{itemize}

Sequence search is a key enabling component to the reuse of metagenomic data deposited in SRA and is necessary for scientists to explore challenging use-cases and ask novel questions within these datasets, which is why metagenomic datasets were chosen as the target for the codeathon.

\subsubsection{Examples of scientific inquiries/discoveries based on metagenomic analysis}
\label{sec:ExamplesOfMetadatAnalysis}
A great example of the power of leveraging multiple public datasets to advance biological knowledge is the discovery of a new virus, a bacteriophage now referred to as CrAssPhage~\cite{dutilh2014highly} 
This virus was discovered by combining data from multiple metagenomic samples generated by the Human Microbiome Project ~\cite{turnbaugh2007human}. This computational discovery was validated experimentally, and CrAssPhage was shown to be a virus that infects bacteria from the genus \textit{Bacteroides} and was found to be highly prevalent in human stool samples, to an extent that led to a proposal for its use as a diagnostic for fecal contamination in environmental samples~\cite{park2020crassphage}.

Furthermore, gene catalogs ~\cite{qin2010human,li2014integrated,ma2020comprehensive}
developed by clustering gene sequences across large numbers of metagenomic samples, have tremendously expanded the scope of known bacterial genes, and led to the development of binning algorithms able to reconstruct novel genomes from metagenomic data ~\cite{nielsen2014identification,parks2017recovery}.


\section{Current approaches to sequence search}
\label{sec:SeqSearchApproches}

Sequence-based searches attempt to match a sequence string query to a target sequence within the search set. Queries can vary in sequence length as well as in terms of the identity between query and intended search targets. The level of sequence divergence that needs to be tolerated by a search algorithm depends on use cases that range from identifying samples that contain near-exact matches to a query sequence (read mapping or variant detection)  to identifying sequences
that are related to the query but are highly diverged due to accumulated mutations.
Sequence search has historically been dominated by sequence alignment, which can be performed with (a) sequence-to-sequence alignment methods (BLAST, SW-align, NW-align); (b) sequence-to-profile alignments methods ~\cite{krogh1994hidden} for increased sensitivity using profiles learned from related family members; 
or (c) profile-to-profile alignment \cite{soding2005protein,wu2008muster}.
There also exist several approaches that use heuristic methods to accelerate sequence search using k-mer Minhash-based probabilistic data-structures 
to estimate genomic distances (instead of computing exact alignments) especially when the goal is to find highly similar sequence matches. K-mer indexing methods serve as effective proxies for exact and approximate sequence searches and can be used effectively 
to reduce memory and computational resource requirements for genomic data analysis \cite{karasikov2020metagraph, solomon2016fast, pandey2018mantis, yu2018seqothello}.
In this section, we focus on tools that implement indexing and/or alignment based strategies. 

\subsection{Alignment-based searches}
\label{sec:AlignmentBased}

Sequence alignment is a method that arranges the letters of biological sequences to highlight their relationships, and can be used to recognize non-random similarity between a query and target sequence.
Alignment is irreplaceable in many aspects of today's biology, such as the annotation of conserved protein domains and motifs, tracking phenotype-related sequence polymorphisms, reconstruction of ancestral DNA sequences, determining the rate of sequence evolution, and homology-based modeling of three-dimensional protein structures, all of which require highly sensitive matching.

The workhorse of sequence alignment for large database search has long been BLAST~\cite{altschul1990basic}.
In the $>$30 years since BLAST was developed, alignment-based database search has been the target of intense algorithmic design for increased speed. Highly-accelerated algorithms such as MMseqs2~\cite{steinegger2017mmseqs2} and DIAMOND ~\cite{buchfink2021sensitive} and LAST ~\cite{kielbasa2011adaptive} are designed for increased speed motivated by increasing scale of sequence repositories.
These speed gains are of vital importance, but come at the cost of loss in sensitivity. Those sensitivity gaps are particularly relevant in the context of microbial community datasets, where annotation efforts often fail to identify large fractions (and in many cases, the majority) of putative proteins. 

Orthogonal to algorithmic speed gains, 
superior sequence search sensitivity~\cite{karplus1998hidden,krause2024sensitive} has been achieved using representations of sequence families by position-specific scoring matrix (PSSM, as in PSI-BLAST~\cite{altschul1997gapped} and MMseqs2~\cite{steinegger2017mmseqs2}) or profile hidden Markov model (pHMM, as in SAM~\cite{karplus1998hidden}, HMMER~\cite{finn2011hmmer}, and nail~\cite{roddy2024nail}), albeit with reduced speed relative to the fastest tools. 
These trade-offs are of immense interest to researchers annotating genomes and the protein sets that they encode.

\subsection{Index-based searches}
\label{sec:IndexBased}

Performing alignment-based sequence-level searches on large collections of SRA and WGS data can quickly become impractical due to size of the databases. For instance, given a query sequence, searching across all metagenomes in SRA would entail a search against >6.5 million accessions.  Such searches are hard to scale due to the use of complex data-structures and significant memory and time footprint. 
As a result, users will have to limit their searches based on attributes such as taxonomy, metadata, sequencing platform etc. Alternatively, several index-based approaches break the search space down to short non-overlapping sequence strings of a fixed length called “k-mers”. These k-mers are then associated either directly with the sample from which they originated or with organisms or other biological markers in the sample and stored in an index. The query sequence is broken down into overlapping k-mers of the same length that are then used to “look up” samples that include matching k-mers, organisms, or biological markers. Notable examples can be grouped into three categories: (a) methods that utilize sketching techniques using one or more hashes to summarize input data like Mash~\cite{ondov2016mash,ondov2019mash}, Sourmash/Branchwater~\cite{pierce2019large,irber2022sourmash}, Kraken~\cite{wood2014kraken,wood2019improved}, STAT~\cite{katz2021stat}; (b)  methods that utilize bloom filter-based data structures to allow for approximate membership queries as seen in BIGSI (BItsliced Genomic Signature Index)~\cite{bradley2019ultrafast}, COBS~\cite{bingmann2019cobs}; (c) approaches that utilize exact representations of annotated de Bruijn graphs (colored de Bruijn) in addition to serving as k-mer indexes as seen in Mantis~\cite{almodaresi2020efficient}, Bifrost~\cite{holley2020bifrost}, Raptor~\cite{seiler2021raptor}, MetaGraph~\cite{karasikov2020metagraph}. More recently, promising approaches that utilize new data-structures like the Hierarchical Interleaved Bloom Filter (HIBF)~\cite{mehringer2023hierarchical} can achieve similar compression, handle unbalanced size distributions and markedly speed up query response time while optimizing space consumption. Evidenced by the need for faster approaches, a new software suite at NCBI, Pebblescout ~\cite{shiryev2024indexing} allows real-time searches across eight databases that index over 3.7 petabases. It can execute searches for query sequences of wide range of lengths and can reduce the search space down to select read-sets of interest, thereby substantially reducing the effort for downstream analysis.

\subsection{Alignment-based vs Index-based comparison}
\label{sec:compareSeqMethods}

Index and alignment methods each have their own advantages and disadvantages:
A key observation concerning the core difference between the two approaches, is that the output of alignment-based approaches provide specific indicators to the source of inferred relationship and conserved regions i.e. input sequences are preserved and may be fully recovered from the output thereby preserving the biological context. Alignment-based methods are more sensitive in identifying relationships between sequences differentiated by many mutations. Although alignment methods have substantially improved for speed over the last 30 years, 
they still lack the scalability needed for petabase-scale sequence search. 
In comparison, index-based approaches are orders of magnitude faster but often lack the ability for sensitive alignment especially for diverged sequences. Indexes used to support search can be very large, but there are approaches to reduce size (min-hash, sequential indexes, bloom filters). Index-based approaches require the index to grow incrementally as new sequences (assembled or unprocessed) are deposited. Depending on the organism, there can be enormous diversity (eg. viruses) which could limit scaling the index and impact the index construction time (although a one-time cost). Index-based and alignment-based methodologies are not mutually exclusive. Prior to constructing alignments, most modern sequence alignment tools perform some kind of index-based search to narrow down the set of comparisons subject to full-cost sequence alignment algorithms. 
A focused community-based approach is crucial to identifying a path toward bridging this evident gap and to demonstrate the practical feasibility of indexing the whole SRA and execute petabase-scale query operations.

\section{Challenges for large-scale sequence search}
\label{sec:SeqSearchChallenges}

Data repositories are growing constantly, so as soon as an index is created, it is out of date. Similarly, as the repository’s size increases, alignment-based approaches become too expensive. Although there are fundamental limitations to each sequence search approach explored in the codeathon, these methodologies represent the best tools currently available to address the problem of large-scale search across the SRA corpus. 
A scientist conducts a search for data because they need it to support an experiment or inquiry. Different scientific questions may require different indexes of the sequence data or flexible workflows to set up a series of steps that filter and refine their search. The concept of search must be expanded to include not just 
queries optimized for exploring metadata, but also fast analysis tools that allow a user to reconfigure data structures on the fly to look for relationships. Refinement of search-based workflows yields insights into the strengths and weaknesses of different components, and can pinpoint where resources should be invested. 
Working directly with the raw reads, while challenging, avoids potential issues created by inconsistent data processing. For example, large fractions of metagenomic datasets do not assemble into contigs, so a comprehensive search of information from those samples must be done at the read level. 
Exploration of sequence data is amenable to a hybrid indexing and alignment approach where the inexpensive indexes can be calculated and used to reduce the search space for the more costly and sensitive alignment methods. 

\paragraph{Challenges with scale:}

The sheer size of the growing SRA corpus presents several challenges to both index and alignment methodologies. Indexes must scale to accommodate tens of millions of samples and support efficient iterative updates and insertion of new data. Patterns of interest may also be spread across multiple reads in a sample (no contiguous substring), preventing successful identification of the complete string, especially when searching across millions of unassembled SRA read sets. Though assembled reads offer some advantages, some data is lost compared to the corresponding unassembled reads in original sample. Such reference representations reduce the sequence content to a single string which is lossy and masks copy number (e.g., gene expression) and nucleotide variations in the original samples. Even though the cloud may offer computational solutions to implementing search strategies, cloud-native tools are needed to bridge download and format transformation time performance bottlenecks when working with SRA data. Tools are required that can perform these operations at scale, in parallel, leveraging a distributed cloud native environment.

\begin{figure}[t]
        \centering
        \includegraphics[width=0.8\textwidth]
        {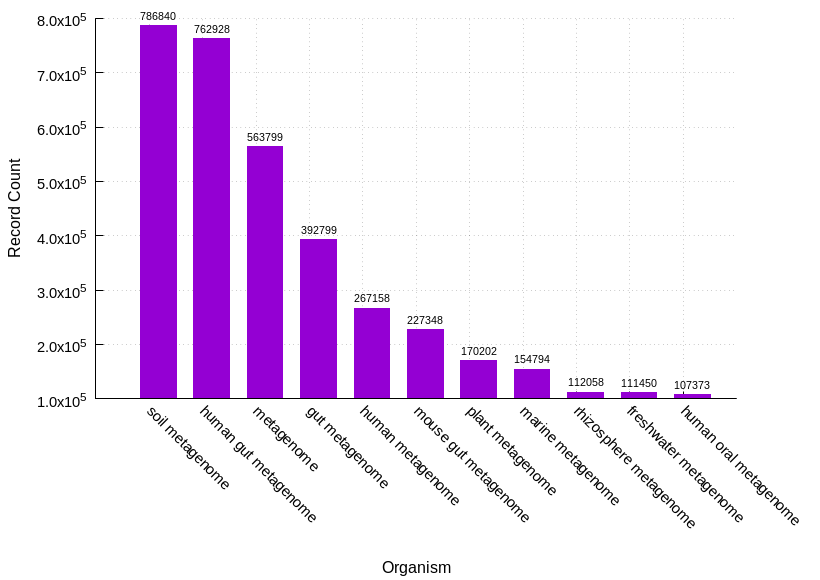}
        \caption{Top metagenomic accession record counts classified based on type `scientific organism name' as of Dec 2024.}
        \label{fig:srametagenomebyorg}
\end{figure}

\paragraph{Challenges with metadata:}

The inherent diversity of metagenomic datasets makes metadata searches difficult. As of December 2024, the metagenomic (and metatranscriptomic) subset of SRA is comprised of over 6.5M samples. Fig. 4 represents the top organism categories (in record count) of samples classified as metagenomic. As seen the organism names for these samples is not always informative, and it is difficult to identify samples of interest based on these names alone. Moreover, metadata alone does not provide insights into the overall taxonomic content within a given sample, potential contamination, or erroneous labeling of the initial sample. Furthermore, SRA metadata include many synonyms, spelling variants, and references to outside sources of information. Manual annotation of the data remains intractable due to the large number of samples in the archive. For these reasons, it has been difficult to perform large-scale analyses that study the relationships among diverse diseases, tissues, and cell types present in SRA.

\section{The Metagenomics Benchmarking Codeathon}
\label{sec:psss-codeathon}

\begin{figure}
   \begin{minipage}[b]{.55\linewidth}
        \centering
        \includegraphics[width=0.95\textwidth]{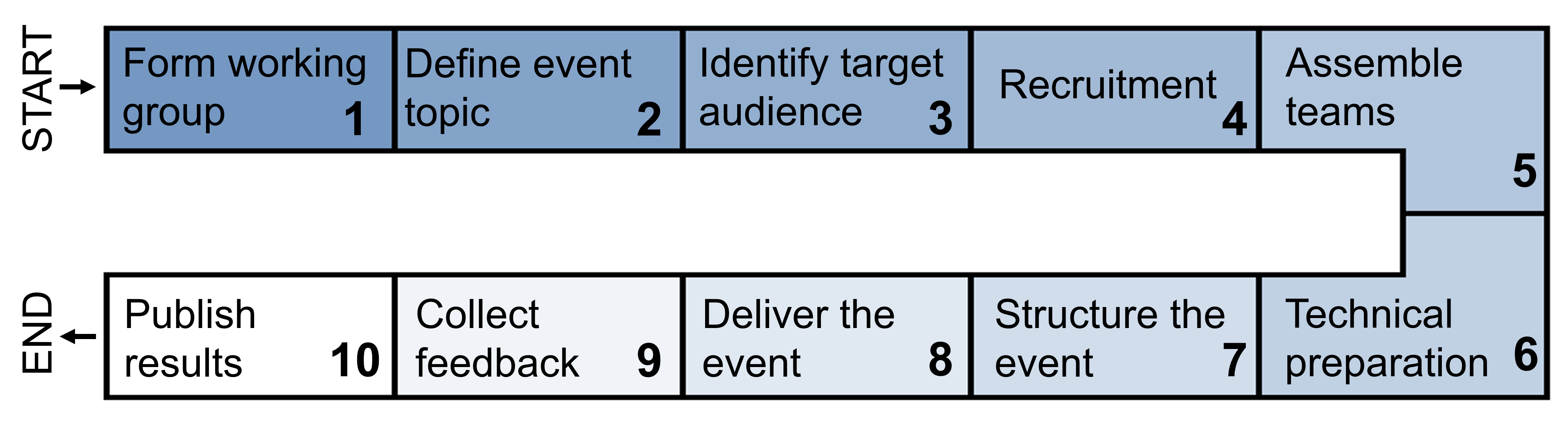}
        \captionof{figure}{Codeathon planning overview.}
        \label{fig:codeathonsteps}
    \end{minipage}\hfill
     \begin{minipage}[b]{.45\linewidth}
        \centering
        \begin{tabular}{|l|l|}
            \hline
            NIH               & 12\% \\ \hline
            DOE National Laboratories & 24\% \\ \hline
            Industry          & 12\% \\ \hline
            Academia          & 52\% \\ \hline
        \end{tabular}
        \captionof{table}{Participant affiliation}
        \label{partipantaffiliation}
     \end{minipage}
\end{figure}

The codeathon was hosted virtually by the NIH Office of Data Science Strategy, NCBI, and DOE’s BER. It attracted experts from U.S Department of Energy national laboratories, scientists and early career researchers from multiple research institutions, faculty and students from universities across the world. In total, 40 participants worked together to develop benchmarking approaches to address challenges in conducting large-scale analyses of metagenomic data. Table~\ref{partipantaffiliation} and Fig. \ref{fig:codeathonsteps} represent the breakdown of participant affiliation and steps to plan and deliver the codeathon respectively. 

\begin{figure}[t]
        \centering
        \includegraphics[width=0.95\textwidth]{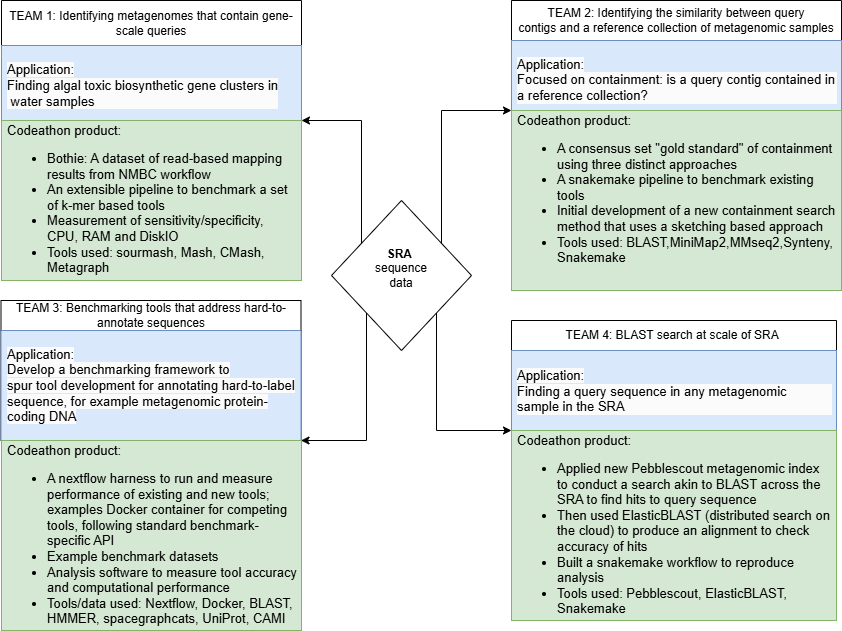}
        \caption{Codeathon Team Projects overview.}
        \label{fig:teamprojects}
\end{figure}

Codeathon participants were split into four teams. The team leaders attended an online, pre-codeathon event to orient them to the data and working in a cloud environment. This also allowed them to further refine the scope of the problem and the focus for each of the
various groups, described  in Fig. \ref{fig:teamprojects}) and generate the following proof-of-concept or early-stage solutions:
\begin{enumerate}
    \item a pipeline used for the identification of metagenomic samples with user-provided long sequence queries,
    \item a gold-standard dataset and pipeline to benchmark contig containments,
    \item a framework evaluating the sensitivity and accuracy of sequence alignment methods on datasets containing challenging inputs, and
    \item a pipeline to combine an experimental SRA sequence index with BLAST.
\end{enumerate}

\subsection{Data Preparation}
\label{psssdataprep}

For the purpose of the codeathon, a range of samples were assimilated from across different complexities. \url{https://portal.nersc.gov/cfs/m3408/PBSequenceSearch/} contains information for metagenome datasets from three different environments (human gut, marine, and soil). The main directory contains a tab-separated file (data.tsv) describing the datasets and outputs, and contains all the relevant dataset identifiers. The raw reads can be obtained from NCBI SRA. The site hosted by the DOE National Energy Research Scientific Computing Facility (NERSC), contains all the analysis products: QA/QC, reads-based analysis, assembled and annotated data (along with MAG's obtained from binning) obtained from National Microbiome Data Collaborative (NMDC) workflows~\cite{nmdcworkflows} to generate the 55 gut microbiome~\cite{nihhumanmicrob}, 33 marine~\cite{larkin2021high} and 1 soil sample(s)~\cite{soilsamplemetadata}. 

\section{Team Projects}
\label{sec:team_projects}

\subsection{Team 1: Identifying metagenomic samples with user-provided long queries}
\label{sec:team1}

\subsubsection{Background}

Microbial environments are being explored on a genomic level like never before, with hundreds of thousands of metagenome datasets now available in SRA. Each metagenome is a rich snapshot of its environment and may contain both known and unknown genes and genomes, including novel unculturable microbes. A critical first step to leveraging these data is the fast and accurate detection of query genes and genomes within each metagenome. For example, identifying metagenomes that contain specific microbes can help explore their potential to produce compounds such as novel antibiotics, medicines, or toxins, or may help elucidate their role in geochemical cycling (e.g. nitrogen and carbon cycle).

Our group focused on the evaluation of lightweight, k-mer based search approaches that have potential as high-throughput screening methods. The strategy was to start by implementing benchmarks for several k-mer based tools for sequence search, parsing the results from provided metagenome datasets for comparison and develop datasets that challenge k-mer based tools.  In particular, we focused on four tools: CMash, Mash, sourmash, and Metagraph, which have all previously been demonstrated to allow genome-scale query detection \cite{irber2024sourmash, ondov2019mash, liu2022cmash, karasikov2020metagraph}. Mash, Cmash, and sourmash all use sketching techniques for faster search and allow k-mer containment searches for accurate sequence similarity comparisons between datasets of different sizes. While Metagraph does not use sketching techniques, its indexing scheme allows fast k-mer search and comparison at a similar scale. Finally, in addition to DNA k-mer searches with all tools, we evaluated protein k-mer searches with sourmash to evaluate capacity for more sensitive search across nucleotide substitutions.

\subsubsection{Datasets}

To evaluate detection accuracy with sub-genome scale queries, we developed a test focusing on toxin biosynthesis gene clusters in environmental water samples, beginning with a test dataset sequenced from a toxic harmful algal bloom in the Saint Lucie Estuary. Since the goal is to screen many metagenomes for a gene cluster of interest, we indexed the Microcystin gene cluster (GenBank AF183401.1) as a database for fast search. This cluster is 64,534 bp in length.

\subsubsection{Codeathon product}

While many tools can conduct sequence analysis and classification of metagenomes, few high-throughput screening methods have been developed for sub-genome level screening at SRA scale. We developed “bothie,” \footnote{\url{https://github.com/NCBI-Codeathons/bothie}} a pipeline that can be used to detect user-provided long queries such as viral, bacterial, fungal, and micro-eukaryotic genomes or biosynthetic gene clusters in metagenomic samples. This workflow evaluates sequence detection using a number of k-mer based search tools, with BLAST-based comparisons used as likely groundtruth. For each tool, we developed a small Snakemake workflow to execute software installation, index the sequence of interest, and search with one or more metagenomes \cite{molder2021sustainable}. For k-mer sketching methods, “detection” was defined as containment-based similarity above the default threshold of each tool. The bothie workflow then links all tool workflows via a python entry point, enabling users to query all tools in a single run. This structure is extensible, allowing addition of new tools for comparison, with all methods for each tool defined within tool-specific workflow files. 

\subsubsection{Results}
\begin{itemize}
    \item Identifying toxin biosynthesis gene clusters in environmental water samples.
    \item Query - Microcystin BGC GenBank AF183401.1 : 64,534 bp
    \item Reference - Toxic Harmful Algal Bloom in the Saint Lucie Estuary: metagenome size 3,448,844 bp
    \item Method - sourmash sketch dna and sketch translate
    \item Result - Found biosynthetic gene clusters in water sample with 56\% similarity
    \item Conclusion - Positive control works. Approach can be implemented in other water samples to screen for Microcystin.

\end{itemize}

\subsubsection{Future Work}

The bothie workflow provides an extensible framework for evaluating the detection accuracy and resource utilization of candidate tools on metagenome samples. While during the codeathon we focused on detection of the microcystin biosynthetic gene cluster as a positive control, future work should focus on developing a series of groundtruth datasets that can be used to assess detection accuracy and limits. For the bothie workflow, work could focus on integrating additional high-throughput screening tools for search and testing with additional queries, including smaller gene-level queries. In particular, the team is interested in testing search for eukaryotic sequence, which has not been a strong focus for metagenomic search tools. 
\subsection{Team 2: Identifying the similarity between query contigs and a reference collection of metagenomic samples}
\label{sec:team2}

\subsubsection{Background}
\label{team2:background}

Our group's project lay focus on evaluating the performance and accuracy of tools for identifying the similarity between one or more query contigs and a "reference" collection of metagenomic samples. 
In the interest of time, the focus was specifically on containment relationships. For the purposes of this application, we define a "match" between two contigs as: 
given contig A of length $len_{A}$ and B of length $len_{B}$, we say that A and B have a containment relationship if an alignment exists between A and B that has higher than 95\% identity within the aligned region, and that covers more than 95\% of the length of the shorter contig. 
Additional cases of containment relationships have been provided here~\cite{team2wiki}.
To tackle the problem we divided it into three sub-tasks:
\begin{itemize}
    \item Find what current solutions exist for this problem. Are there avenues to develop new solutions to improve the performance? 
    \item Find ways to benchmark contig containments 
    using existing approaches by (a) generating a "gold standard" (groundtruth) dataset for testing purposes, and (b) identifying evaluation metrics and (c) develop new benchmarking tools.
    \item Build a Snakemake pipeline with query and reference data as input and produce a benchmarking report as output with containments. 
\end{itemize}
We explored three existing methods that are alignment-based (BLAST, MMseq2, Minimap2), two that are hash-based (dashing, MashMap), one is based on synteny maps (shared ordered information) using the R package DECIPHER. 
Table \ref{tab:align_methods}, tabulates the pros and cons of each approach.

\begin{table}[H]
\caption{Tools for identifying similarity}
\begin{tabular}{|l|l|l|}
\hline
\multicolumn{1}{|c|}{\textbf{Tool}} & \multicolumn{1}{c|}{\textbf{Pros}}                                                                                                                                            & \multicolumn{1}{c|}{\textbf{Cons}}                                                                                                                                    \\ \hline
BLAST\cite{ye2006blast}                              & Widely-used tool that is broadly available                                                                                                                                    & May be slow for large datasets                                                                                                                                       \\ \hline
MMseq2\cite{steinegger2017mmseqs2}                              & Faster than BLAST without sacrificing accuracy                                                                                                                                & \begin{tabular}[c]{@{}l@{}}Since it computes full alignments, \\ may be slow for large datasets\end{tabular}                                                          \\ \hline
Minimap2\cite{li2018minimap2}                            & \begin{tabular}[c]{@{}l@{}}chaining based accurate and efficient alignment, \\ specifically designed for long genomic segments\end{tabular}                                   & Slower for large datasets                                                                                                                                             \\ \hline
dashing\cite{baker2019dashing}                             & \begin{tabular}[c]{@{}l@{}}Hash-based approach based on a HyperLogLog \\ sketch, much faster than alignment-based methods\end{tabular}                                         & \begin{tabular}[c]{@{}l@{}}Accuracy may be limited \\(as alignments are not computed), \\does not estimate position of a match\end{tabular}                            \\ \hline
Synteny\cite{wright2016using}                             & \begin{tabular}[c]{@{}l@{}}Synteny maps generated from unique shared \\ k-mers, High precision - finds shared information \\ in both coding and non-coding space\end{tabular} & \begin{tabular}[c]{@{}l@{}}Slow (relative to other tools for this usage), \\limited by R's access to memory, \\ recall sacrificed in favor of precision\end{tabular} \\ \hline
MashMap\cite{jain2017fast}                             & Alignment free hash-based approach.                                                                                                                                           & \begin{tabular}[c]{@{}l@{}}Accuracy is limited as alignments \\are not computed\end{tabular}                                                                                                                   \\ \hline
\end{tabular}
\label{tab:align_methods}
\end{table}

\subsubsection{Datasets}

\begin{figure}[H]%
    \centering
    \subfloat[\centering Containments obtained for BLAST, MMSeq2 and Synteny maps]
    {{\includegraphics[width=6cm]{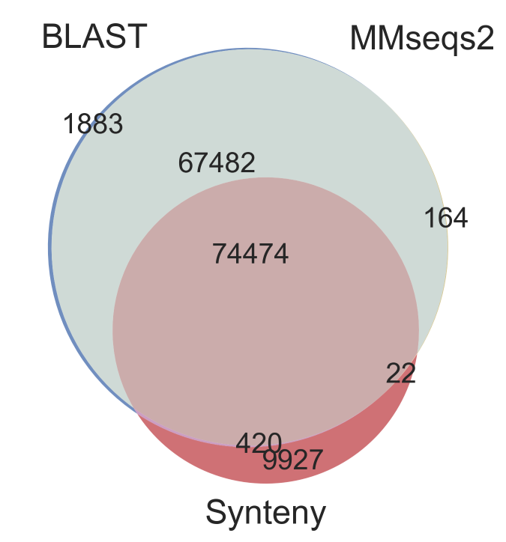} }}%
    \qquad
    \subfloat[\centering Performance of the Snakemake pipeline obtained for BLAST and MMSeq2]
    {{\includegraphics[width=7cm]{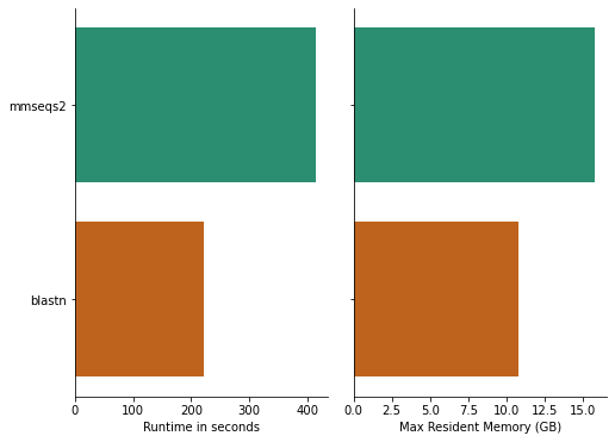} }}%
    \caption{Preliminary results obtained for the marine metagenome dataset}%
    \label{fig:marinemetaresults}%
\end{figure}

\paragraph{Marine Metagenomes}: 30 Ocean microbiomes from NMDC (accessions under ocean metagenome dataset in \url{https://portal.nersc.gov/cfs/m3408/PBSequenceSearch/}). We defined the sample nmdc:mga04781 as the query sample and the remaining samples as the reference database. Only contigs >= 500 bp were used in the searches. Containments were identified using BLAST, mmseqs2, and synteny maps, wherein we searched one of the marine metagenomes against the other 29. The results of using BLASTn for the search were treated as the gold standard.
Our evaluation using the marine metagenome dataset (as seen in Fig.  \ref{fig:marinemetaresults}) found a total of 155,372 containments. BLAST and MMseq2 agreed on a larger fraction (>141k containments) whereas the Synteny-based method found fewer containments in common ~75k.

\paragraph{Marine + Human Gut Metagenomes}: 30 Ocean microbiomes and 55 human gut microbiomes (accessions in Appendices \ref{appendix_1}\&\ref{appendix_2}).  We defined the sample nmdc:mga04781 as the query sample and the remaining samples as the reference database. Only contigs >= 500 bp were used in the searches. Containments were identified using BLAST, MMseqs2, and synteny maps. The results of using BLASTn for the search were treated as the gold standard.

\paragraph{Stool Metagenomes}: Stool microbiomes from \cite{sharon2013time} (accessions in Appendix \ref{appendix_3}) represent a time series of 17 fecal samples from a single premature infant. Only contigs >= 500 bp were used in the searches. Containments were identified using all vs all alignments to obtain precision and recall values on BLAST, Dashing, MiniMap2 and MashMap. The results of using BLASTn for the search were treated as the gold standard.

\subsubsection{Codeathon product}

We have generated code that constructs a gold standard for a given dataset, code that evaluates a tool with respect to the gold standard, and also provided several implementations of tools that solve the containment problem.  In addition, we have generated gold standards for two datasets, one comprising ocean metagenomes, and the other comprising human stool metagenomes.

A ground truth set of containments was computed using BLAST. MMseq2 and Synteny-based methods were subsequently added to the Snakemake pipeline, allowing each tool's output to be evaluated against this standard. Each record was classified as a true positive (TP, indicating truth and tool agree a match exists), true negative (TN, indicating that truth and tool agree there is no match), false positive (FP, indicating the tool reported a match that is not found in the truth), and false negative (FN, indicating the tool did not report a match found in the truth).  In addition we computed the precision(\(\frac{TP}{TP + FP}\), correct predictions as a fraction of all predictions) and recall (\(\frac{TP}{TP+FN}\)) correct predictions as a fraction of all true matches).  The recall was further stratified by the percent identity of matches and length of contigs as represented in the truth. The precision was further stratified by the confidence (when available) reported by the tool as well as by the length of the contigs.

In addition, we captured information about the run time and memory usage of the tool being benchmarked using both Snakemake logs and direct measurement in the workflow.

We also explored an additional novel approach to the problem that is sketching based. Most existing sketch based methods do not retain order information, whereas our new method sketched the reference database using a sketch that retains the order and maintained bins of minimizers (to sketches) that contained the minimizer value and computed (within each bin) the minimum hamming distance between any query sequence and the reference sequences. The sequences filtered based on the hamming distance were subsequently filtered based on the average nucleotide identity. Going forward we would hope to expand this approach to build a more efficient index using colored minimizer de Bruijn graph representations within each bin that may enable us to answer queries in nearly constant time. We would also like to explore ways to extend this approach to other kinds of contig-contig relationships.
The implementation of the work conducted by our team can be found within the repository: \url{https://github.com/NCBI-Codeathons/psss-team2}.

\subsubsection{Future Work}

The initial focus of the codeathon was on the relatively constrained set of matches between contigs that represent containment relationships (one contig is fully contained in another). Going forward our future plans entail:

\begin{itemize}
    \item We plan to expand the analysis beyond containment to include a general definition of "substantial match" between metagenomic contigs (taking into account overlaps,insertions). Work is needed to expand the definition and tie it to specific analytic use-cases, and adapt the existing code to take the new definition into account.
    \item  We also wish to add additional tools to the Snakemake pipeline or make it easier for others to submit new tools to the pipeline as well.
    \item Finally we would like to improve our Snakemake pipeline to scale to larger datasets.
\end{itemize}  

\subsection{Team 3: Benchmarking tools addressing hard-to-annotate sequence}
\label{sec:team3}

\subsubsection{Background}

While petabyte-scale search naturally emphasizes speed above all else, the focus of our group was on ensuring that tool developers and users can understand the sensitivity trade-offs that come with methods for increasing speed. 
In a way, it is straightforward to develop a benchmark for sequence annotation tools: gather a bunch of positive controls (query/target pairs that share trusted homology), supplement them with decoys (candidates for annotation that are known not to be related), and select some measure that enables evaluation of how well the software distinguishes the positives from the decoys. But this process involves innumerable decisions regarding selection of positive controls, design of decoy set, exploration of subtleties such as overextension~\cite{frith2008whole,gonzalez2010homologous,hubley2016dfam}, benchmark size, metric selection, and the potential for unintended consequences of design decisions~\cite{glidden2024match}. 

Despite universal interest, evaluation of database search efficacy has generally fallen to the tool developer, with each developer (re-)inventing assessment approaches and only partially exploring the full range of challenges presented to their software in real-world use cases. Without a standard, results across papers can be difficult to compare. We sought to use the short codeathon to lay the groundwork for a rigorous and community-supported suite of annotation benchmarks, with an emphasis on evaluating the performance of alignment/annotation methods faced with difficult-to-label (meta-)genomic sequence. Well-designed and community-supported benchmarks reduce impediments to attracting new researchers to the domain while also enabling principled evaluation of alternative methods/software for annotation tasks across multiple perspectives. Specifically, we:

\begin{itemize}
    \item Developed a small number of example benchmark datasets (and corresponding analysis scripts) representing situations in which accurate annotation is difficult.
    \item Designed the benchmark generation scripts to allow flexible creation of datasets with varying levels of difficulty.
    \item Developed a general test harness that makes it relatively easy for benchmark users (tools developers and others who wish to assess performance) to run tests on large-scale cluster/cloud resources without needing to write code to control those resources. 
    \item Furthermore, we aimed to design the benchmark harness in a way that allows it to serve as a low-barrier framework that others can use to develop future benchmarks.

\end{itemize}

Rather than design our benchmarking strategies from scratch, we were guided by existing examples of well-crafted benchmarks for challenging search scenarios. In the profmark benchmark~\cite{eddy2011accelerated}, all sequences in Pfam~\cite{el2019pfam} seed alignments are considered to be true matches, and decoys are generated by shuffling sequences sampled from Uniprot~\cite{mcgarvey2019uniprot}. An alternative approach is to establish a collection of ``true'' positives based on  Uniprot sequences sharing folds defined by SCOP~\cite{andreeva2007data}, with decoys either defined as sequence pairs from different SCOPe folds~\cite{buchfink2021sensitive} or by reversing Uniprot-sourced protein sequences~\cite{steinegger2017mmseqs2}. These benchmarks for protein-to-protein search methods cover many common annotation scenarios, and can be supplemented by benchmarks for annotation of DNA-DNA search~\cite{wheeler2013nhmmer} (such as when seeking genomic instances of transposable elements~\cite{wheeler2012dfam} or functional RNA~\cite{nawrocki2009infernal}). A less-explored  benchmarking context is that of translated search: seeking protein-coding DNA regions that correspond to one or more target protein sequences or profiles.

\subsubsection{Codeathon product}

Considering the PSSS focus on searching against SRA, we settled on three translated search benchmarks that would be (i) relatively simple to generate under the codeathon time constraints, (ii) probably useful for developers wishing to develop tools that annotate hard-to-label inputs, and (iii) lay the groundwork for extended variants of the benchmark that incorporate additional challenges, such as splicing (which can cause fragmentation in eukaryotic annotations) and frameshifted differences between query protein and target coding DNA (due to anything from sequencing error to pseudogenization to programmed ribosomal frameshifting). 
We captured each pilot benchmark implementation in folders within the repository \url{https://github.com/NCBI-Codeathons/psss-team3-hard-annotation}:

\begin{itemize}
    \item protein domains embedded within DNA decoys (based on the approach used in profmark) 
    \item incomplete protein domains (abbreviated contigs, modified from the above benchmark)
    \item annotating reads that resist assembly - using mock community reads from CAMI (Bremges and McHardy 2018), build assembly graph, identify reads (and associated sub-graph) that fail to be extracted into an assembled contig, and determine how much annotation coverage is lost relative to the source (known) microbial genomes.

\end{itemize}

\paragraph{Benchmark metrics (high-level):}

Some measures of efficacy are essentially universal among annotation benchmarks: sensitivity (fraction of planted positives that are recovered by the tool), false labeling rate (sometimes reported as a false discovery rate but more often presented in contrast to the sensitivity, for example to produce ROC curves or compute the area under such a curve), and computational performance (run time and memory requirements). 

In our pilot effort, we sought to lay the groundwork for measuring other important aspect sof annotation, including: alignment accuracy, boundary detection, and correct handling of confounding sequences such as those subject to sequencing error or containing repetitive sequence.

\paragraph{A generic benchmark harness:}

A benchmark includes both the actual data used in tool evaluation and supporting scripts used to perform evaluation and produce summary statistics/figures.

During (and shortly after) this codeathon, we also developed a pilot implementation of a harness for running the key benchmarking across multiple platforms (local, cluster, cloud) in a consistent and straightforward manner. This harness, code-named BAGEL~\cite{BAGEL}, depends on the Nextflow scripting language~\cite{di2017nextflow}; once a benchmark is created, a new tool may be tested in the benchmark's harness by creating a simple docker image that links the tool up to the harness' API.

\subsubsection{Future Work}

This pilot project yielded sample benchmarks and an associated test harness that explored and demonstrated reasonable steps to be taken in building a general benchmark that is not built for evaluation as part of development effort for a single competitor tool. Going forward, there remains much work to be done

\begin{itemize}
    \item Expand harness to full working order. 
    \item Complete the aforementioned benchmarks. 
    \item Expand to new related assessments, particularly (i) impact of sequencing error, (ii) homologous overextension, and (iii) improved CAMI-style assembly graphs, over new CAMI data.
    \item Include full-length proteins, e.g. accounting for challenges caused by domain architecture
    \item Use SCOP folds as true positive dataset
    \item Include challenging benchmark for DNA that doesn’t encode proteins (e.g. ncRNA, transposable elements).

\end{itemize}
\subsection{Team 4: Pipeline to BLAST against SRA}
\label{sec:team4}

\subsubsection{Background}
Searching a query sequence against the entire metagenomic subset of SRA database to produce BLAST alignments has not been done previously on a large-scale, though two new tools recently provide this capability. Pebblescout uses a kmer-based indexing approach to handle the enormous task of sifting through the (metagenomic) SRA database, filtering results to the more manageable level of only those containing positive hits, then Elastic-BLAST leverages cloud computing (either GCP or AWS) to allow large-scale pairwise sequence alignments of the resulting datasets with the original query.

\subsubsection{Datasets and Tools}

\paragraph{Pebblescout:} A preliminary version of the Pebblescout~\cite{shiryev2024indexing} metagenomic index comprised of a subset of SRA as of August 12, 2021. The tool allows users to almost `BLAST against SRA' but in an alignment-free approach. This was limited target database returning search results to a maximum of 2000 read sets (at the time) in the output by default. The search returns a ranked list of accessions with matches to the query sequence, which allows the user to subsequently perform additional computations (alignment or assembly) with a much smaller subset of the entire SRA.

\paragraph{ElasticBLAST:} a cloud-native command line tool~\cite{camacho2023elasticblast} that aligns large volumes of nucleotide and protein
sequences against popular and user-provided BLAST databases. Given a large number of queries (in this project it
ranged up to tens of billions of bases), ElasticBLAST splits the queries into batches (on GCP with Kubernetes,
on AWS with AWS Batch), starts the number of requested machines, performs the alignments in batches across the
machines. Finally, ElasticBLAST shuts down the instances and writes the results to a cloud
bucket (specified by the user).  

Our workflow in Fig. \ref{fig:team4pipeline} generates the alignment by connecting the tools Pebblescout and ElasticBLAST.


\paragraph{Query sequences (versions used at the time of the codeathon):}
\begin{itemize}
    \item SARS-CoV-2: NC\_045512.2 \cite{Sars2} 
    \item Crassphage: NC\_024711.1 
    See Post-conference note added to \cite{team4wiki} 
    \item Bacteriophage T4: NC\_000866.4 \cite{bacteriophage} 
    \item E. coli: NC\_000913.3 \cite{ecoli} 
    \item Ebola: NC\_002549.1 \cite{ebola} 
    \item Klebsiella phage K11: NC\_011043.1 \cite{klebsiella} 
    \item Acidianus filamentous virus 1: NC\_005830.1 \cite{acidianus} 
    \item Ancient caribou feces associated virus: NC\_024907.1 \cite{caribou}  
\end{itemize}


\subsubsection{Codeathon product}

\begin{figure}[t]
        \centering
        \includegraphics[width=0.95\textwidth]{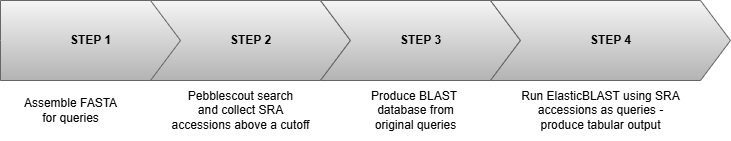}
        \caption{Proposed workflow}
        \label{fig:team4pipeline}
\end{figure}

\begin{figure}[t]
        \centering
        \includegraphics[width=0.4\textwidth]{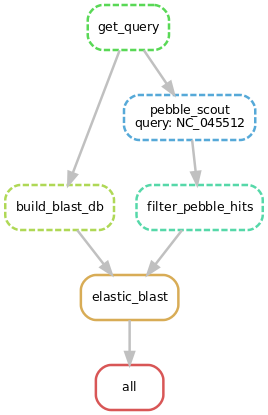}
        \caption{Proposed Snakemake workflow}
        \label{fig:team4SMworkflow}
\end{figure}

Major steps of the pipeline and corresponding Snakemake workflow (as illustrated in Figs. \ref{fig:team4pipeline} and \ref{fig:team4SMworkflow} respectively) are comprised of the following steps for each sequence in the query dataset:
\begin{enumerate}
    \item run Pebblescout search (using the metagenomic index) with the query sequence which in turn returns a list of SRA accessions and corresponding scores that have matches to the query.
    \item filter SRA datasets based on some criteria - discard SRA accessions with sequence read length below a specified threshold, availability in GCP.
    \item build a BLAST target database with the original queries. (note: blastn\_vdb allows to search against SRA run but is not supported by ElasticBLAST).
    \item dump FASTA files from the selected SRA runs and run Elastic-BLAST with the set of filtered SRA datasets as the large set of queries (to run in parallel) against the new (original query) database as the target.
\end{enumerate}

The major bottleneck proved to be the downloading of SRA datasets (identified by Pebblescout, used as query inputs to ElasticBLAST) to the compute instance - the fasterq-dump process (uses multiple threading to download multiple SRA datasets at a time) was incredibly slow, requiring $\sim$20 hrs to handle merely $\sim$400 datasets, and worse was limited by network access rather than CPU speed. Prior to that, Pebblescout could identify several tens of thousands of SRA datasets in mere seconds (a fraction of a minute even), and once completed, ElasticBLAST could also run fairly quickly (1 hour for $\sim$60 billion bases; and even that was mostly shuffling FASTA files around too), but the shuffling of files around makes the current version of this pipeline practically limited to only a few hundred datasets at a time.

One solution to download the target accessions (which would be the input to ElasticBLAST), is to adopt a far scalable parallel approach i.e. using Google cloud function (similar to AWS Lambda). The FASTQ files can be downloaded in parallel across 'n' independent instances and subsequently be converted to the FASTA format in parallel. The time taken to complete the operation would be equivalent to the time taken to complete the download of the largest accession and this dramatically increased the efficiency of our pipeline. Details of the pipeline can be found in the group's repository: \url{https://github.com/NCBI-Codeathons/psss-team4}.

\subsubsection{Future Work}

The pipeline exists now to facilitate the ability of future codeathons to explore this further. Future efforts would actually apply this pipeline for learning, or may continue development to streamline it to further enhance its utility.

\begin{itemize}
\item Replace tools in various steps of pipeline to benchmark against time, output sensitivity/specificity
\item Obtaining FASTA files may impose a substantial overhead, design suitable solutions to address this bottleneck
\item Benchmark Pebblescout against other available solutions
\item Provide a detailed step-by-step guide for the pipeline and benchmark
\end{itemize}

\section{Discussion}
\label{sec:discussion}

With more than 38 petabytes of data and growing, SRA and other public resources provide a vast reservoir of information that can be tapped to accelerate biomedical discovery. Working with this data, codeathon participants applied tools and resources to several large-scale biological questions including but not limited to establish biological relationships between specific genes or organism groups including pathogens and environmental attributes like geography, season, biome, etc. In the future, similar tools can utilize the environmental and temporal diversity of SRA samples to investigate the composition of pangenomes across organisms at strain, species, and genus levels.

The codeathon successfully addressed several key challenges to working with metagenomic data at scale: (a) providing a functional cloud platform in which participants could develop analysis pipelines, (b) rapidly producing benchmarks and reusable software, including product documentation and relevant repository links, (c) evaluating the applicability of current standardized search algorithms to handle the complexities associated with diverse microbial community datasets, (d) exploring the practicality of implementing curated databases/indexes - enabling fast sequence-based searches across very large datasets - to generate a filtered subset of samples for subsequent, more detailed analysis.

As far as community engagement goes, the teams successfully elicited a list of concrete sub-problems and demonstrated that these problems are tractable and the outcomes of the codeathon can lay the groundwork for future PSSS events. A number of users spent significant amount of time and effort transitioning their workflows to the cloud and although they appreciated the value of working at large scale, the transition to using the cloud services and learning the infrastructure was challenging. Workflow management tools (Snakemake, Nextflow) are great for reproducibility but they are not trivial to get working in the cloud. Finally, additional resources will be needed to get users to comfortably work with cloud-native tools rather than downloading the data and working locally. There might be some interest to explore options to assess the tools discussed in this paper for future benchmarking challenges like CAMI (Critical Assessment of Metagenome Interpretation)~\cite{meyer2022critical} as evidenced in future work by one of the teams.

The codeathon was followed by subsequent events organized by DOE JGI and NCBI as part of an effort to train the scientific community in cloud computing and petabyte scale sequence search. One such effort was the workshop ‘Hands-On Petabyte Scale Sequence Search of SRA' supported by the DOE JGI, NCBI, and Berkeley Lab IT, which took place as part of the DOE JGI User Meeting in September 2022. The goal of the workshop was to introduce participants to Elastic-BLAST and sourmash/mastiff/branchwater~\cite{sourmashbranchwater} 
by running these tools in a cloud environment. During the workshop, participants used AWS computing resources to analyze data for 2,631 draft genomes generated using the Tara Oceans microbial metagenomic datasets~\cite{tully2018reconstruction}. When considering a focus for the next event, a round or two of user research is needed to determine what might be of most interest or use to the scientific community. Some outreach is still required to attract participants with other areas of expertise (e.g., math, computer science, statistics).

\section*{Author Contributions}

J.R.B and K.F. conceived of the work. T.B., M.P., R.P., J.S.M., T.J.W.,and T.L.M. served as team leads for each of the groups. A.S. and D.P.R helped facilitate the event and provided support to researchers. All others were involved in performing the work described. P.G. prepared the manuscript with review and editing by others. Those with significant contributions to the manuscript preparation include T.J.W., M.P., R.C. and J.R.B.

\section*{Funding}
This work was supported in part by the National Center for Biotechnology Information of the National Library of Medicine (NLM), National Institutes of Health and the NIH Science and Technology Research Infrastructure for Discovery, Experimentation, and Sustainability initiative (STRIDES).

Part of the work was also supported by U.S. Department of Energy Joint Genome Institute (\url{https://ror.org/04xm1d337}), a DOE Office of Science User Facility, supported by the Office of Science of the U.S. Department of Energy operated under Contract No. DE-AC02-05CH11231.

TJW, JWR, TC, and GRK were supported by NIH NIGMS R01GM132600 and  DOE DE-SC0021216. MH was supported by NIH - R01AI100947. RKMX was supported in part by a Harbor Branch Oceanographic Institute Foundation Indian River Lagoon Graduate Research Fellowship. AIK was supported by the U.S. Department of Energy, Office of Science, Office of Advanced Scientific Computing Research, Department of Energy Computational Science Graduate Fellowship under Award Number DE-SC0020347.

\section*{Acknowledgments}

We would like to thank Rayan Chikhi from Institut Pasteur, Université Paris Cité, Paris France; Minghang Li from Seoul National University, Seoul; Nicholas Cooley from Department of Biomedical Informatics, University of Pittsburgh, Pittsburgh PA, USA; Michael Shaffer from Department of Soil and Crop Sciences, College of Agriculture, Colorado State University, CO USA; Thomas Biondi from Los Alamos National Laboratory, Los Alamos, NM, USA; and Mikhail Karasikov from Department of Computer Science, ETH Zurich, Zurich, Switzerland; for participating in the event and for their valuable contributions to their respective teams during the codeathon, without which this work would not have been possible.
This work was supported in part by the National Center for Biotechnology Information of the National Library of Medicine (NLM), National Institutes of Health and the NIH Science and Technology Research Infrastructure for Discovery, Experimentation, and Sustainability initiative (STRIDES).

\section*{Conflicts of Interest}

The authors declare no conflicts of interest.



\bibliography{references}

\bibliographystyle{abbrv}

\section{Appendix}
\label{sec:appendices}

\subsection{Appendix 1: Marine metagenome data accessions}
\label{appendix_1}

Ocean metagenome dataset used to generate gold standard from \url{https://portal.nersc.gov/cfs/m3408/PBSequenceSearch/}. Additional information can be found in  \url{https://github.com/NCBI-Codeathons/psss-datasets}.

\begin{longtable}{|l|l|l|}
\hline
\multicolumn{1}{|c|}{\textbf{MGA\_ID}} & \multicolumn{1}{c|}{\textbf{SRS}} & \multicolumn{1}{c|}{\textbf{TAXA}} \\ \hline
\endfirsthead
\multicolumn{3}{c}%
{{\bfseries Table \thetable\ continued from previous page}} \\
\hline
\multicolumn{1}{|c|}{\textbf{MGA\_ID}} & \multicolumn{1}{c|}{\textbf{SRS}} & \multicolumn{1}{c|}{\textbf{TAXA}} \\ \hline
\endhead
nmdc:mga02197 & SRS7767240 & marine metagenome \\ \hline
nmdc:mga0qh70 & SRS7229387 & marine metagenome \\ \hline
nmdc:mga0b497 & SRS7760193 & marine metagenome \\ \hline
nmdc:mga0x030 & SRS7229758 & marine metagenome \\ \hline
nmdc:mga0hf70 & SRS7229506 & marine metagenome \\ \hline
nmdc:mga04781 & SRS7772446 & marine metagenome \\ \hline
nmdc:mga0py30 & SRS7229216 & marine metagenome \\ \hline
nmdc:mga0nd80 & SRS7758293 & marine metagenome \\ \hline
nmdc:mga06w20 & SRS7758535 & marine metagenome \\ \hline
nmdc:mga05685 & SRS7229213 & marine metagenome \\ \hline
nmdc:mga06s29 & SRS7229719 & marine metagenome \\ \hline
nmdc:mga0w223 & SRS7229769 & marine metagenome \\ \hline
nmdc:mga04z09 & SRS7768376 & marine metagenome \\ \hline
nmdc:mga03004 & SRS7229635 & marine metagenome \\ \hline
nmdc:mga0f307 & SRS7229291 & marine metagenome \\ \hline
nmdc:mga08w22 & SRS7762365 & marine metagenome \\ \hline
nmdc:mga06y14 & SRS7229835 & marine metagenome \\ \hline
nmdc:mga01293 & SRS7229228 & marine metagenome \\ \hline
nmdc:mga0xa97 & SRS7229394 & marine metagenome \\ \hline
nmdc:mga0km57 & SRS7759567 & marine metagenome \\ \hline
nmdc:mga0d402 & SRS7229180 & marine metagenome \\ \hline
nmdc:mga00p32 & SRS7229785 & marine metagenome \\ \hline
nmdc:mga0w126 & SRS7774498 & marine metagenome \\ \hline
nmdc:mga0az15 & SRS7229336 & marine metagenome \\ \hline
nmdc:mga0nv38 & SRS7229572 & marine metagenome \\ \hline
nmdc:mga0ef67 & SRS7229831 & marine metagenome \\ \hline
nmdc:mga0sr51 & SRS7774881 & marine metagenome \\ \hline
nmdc:mga0hn52 & SRS7773438 & marine metagenome \\ \hline
nmdc:mga00z05 & SRS7229600 & marine metagenome \\ \hline
nmdc:mga0w514 & SRS7229557 & marine metagenome \\ \hline
\end{longtable}

\subsection{Appendix 2: Human gut metagenome data accessions}
\label{appendix_2}

Human gut metagenome dataset used to generate gold standard can be found here: \url{https://portal.nersc.gov/cfs/m3408/PBSequenceSearch/}.

\begin{longtable}{|l|l|l|}
\hline
\multicolumn{1}{|c|}{\textbf{MGA\_ID}} & \multicolumn{1}{c|}{\textbf{SRS}} & \multicolumn{1}{c|}{\textbf{TAXA}} \\ \hline
\endfirsthead
\multicolumn{3}{c}%
{{\bfseries Table \thetable\ continued from previous page}} \\
\hline
\multicolumn{1}{|c|}{\textbf{MGA\_ID}} & \multicolumn{1}{c|}{\textbf{SRS}} & \multicolumn{1}{c|}{\textbf{TAXA}} \\ \hline
\endhead
nmdc:mga08j52                          & SRS4582705                        & human gut metagenome               \\ \hline
nmdc:mga0j795                          & SRS4582624                        & human gut metagenome               \\ \hline
nmdc:mga0f210                          & SRS4582697                        & human gut metagenome               \\ \hline
nmdc:mga0em52                          & SRS4582712                        & human gut metagenome               \\ \hline
nmdc:mga0rj68                          & SRS4582699                        & human gut metagenome               \\ \hline
nmdc:mga0nn56                          & SRS4582673                        & human gut metagenome               \\ \hline
nmdc:mga04684                          & SRS4582634                        & human gut metagenome               \\ \hline
nmdc:mga0s705                          & SRS4582709                        & human gut metagenome               \\ \hline
nmdc:mga0ke75                          & SRS4582660                        & human gut metagenome               \\ \hline
nmdc:mga0jz23                          & SRS4582649                        & human gut metagenome               \\ \hline
nmdc:mga0bd70                          & SRS4582665                        & human gut metagenome               \\ \hline
nmdc:mga0zc93                          & SRS4582707                        & human gut metagenome               \\ \hline
nmdc:mga0cx23                          & SRS4582643                        & human gut metagenome               \\ \hline
nmdc:mga09689                          & SRS4582693                        & human gut metagenome               \\ \hline
nmdc:mga0wr54                          & SRS4582661                        & human gut metagenome               \\ \hline
nmdc:mga03a71                          & SRS4582640                        & human gut metagenome               \\ \hline
nmdc:mga07008                          & SRS4582714                        & human gut metagenome               \\ \hline
nmdc:mga0df66                          & SRS4582668                        & human gut metagenome               \\ \hline
nmdc:mga0m312                          & SRS4582653                        & human gut metagenome               \\ \hline
nmdc:mga0k990                          & SRS4582644                        & human gut metagenome               \\ \hline
nmdc:mga0g987                          & SRS4582627                        & human gut metagenome               \\ \hline
nmdc:mga02294                          & SRS4582647                        & human gut metagenome               \\ \hline
nmdc:mga0f404                          & SRS4582641                        & human gut metagenome               \\ \hline
nmdc:mga0cr38                          & SRS4582670                        & human gut metagenome               \\ \hline
nmdc:mga0fc77                          & SRS4582686                        & human gut metagenome               \\ \hline
nmdc:mga0h988                          & SRS4582691                        & human gut metagenome               \\ \hline
nmdc:mga0dx24                          & SRS4582701                        & human gut metagenome               \\ \hline
nmdc:mga0kv36                          & SRS4582711                        & human gut metagenome               \\ \hline
nmdc:mga06686                          & SRS4582625                        & human gut metagenome               \\ \hline
nmdc:mga03d62                          & SRS4582630                        & human gut metagenome               \\ \hline
nmdc:mga0ja86                          & SRS4582704                        & human gut metagenome               \\ \hline
nmdc:mga00w14                          & SRS4582652                        & human gut metagenome               \\ \hline
nmdc:mga0kf72                          & SRS4582674                        & human gut metagenome               \\ \hline
nmdc:mga0j310                          & SRS4582669                        & human gut metagenome               \\ \hline
nmdc:mga00f53                          & SRS5862060                        & human gut metagenome               \\ \hline
nmdc:mga0rn59                          & SRS4582639                        & human gut metagenome               \\ \hline
nmdc:mga0qd82                          & SRS4582626                        & human gut metagenome               \\ \hline
nmdc:mga0rz29                          & SRS4582645                        & human gut metagenome               \\ \hline
nmdc:mga08f61                          & SRS4582695                        & human gut metagenome               \\ \hline
nmdc:mga0md79                          & SRS4582715                        & human gut metagenome               \\ \hline
nmdc:mga0nc83                          & SRS4582651                        & human gut metagenome               \\ \hline
nmdc:mga0rq53                          & SRS4582646                        & human gut metagenome               \\ \hline
nmdc:mga0h309                          & SRS4582655                        & human gut metagenome               \\ \hline
nmdc:mga01r27                          & SRS4582637                        & human gut metagenome               \\ \hline
nmdc:mga0aq39                          & SRS4582681                        & human gut metagenome               \\ \hline
nmdc:mga04490                          & SRS4582658                        & human gut metagenome               \\ \hline
nmdc:mga0jt38                          & SRS4582682                        & human gut metagenome               \\ \hline
nmdc:mga0sm63                          & SRS4582700                        & human gut metagenome               \\ \hline
nmdc:mga0zg81                          & SRS4582628                        & human gut metagenome               \\ \hline
nmdc:mga0x224                          & SRS4582680                        & human gut metagenome               \\ \hline
nmdc:mga0dp45                          & SRS4582623                        & human gut metagenome               \\ \hline
nmdc:mga08v25                          & SRS4582710                        & human gut metagenome               \\ \hline
nmdc:mga0qa91                          & SRS4582662                        & human gut metagenome               \\ \hline
nmdc:mga0ky27                          & SRS4582703                        & human gut metagenome               \\ \hline
nmdc:mga0x709                          & SRS4582642                        & human gut metagenome               \\ \hline
\end{longtable}

\subsection{Appendix 3: Sharon et al human gut metagenome data accessions}
\label{appendix_3}

Human gut metagenome dataset used to generate gold standard from Sharon et al.~\cite{sharon2013time}.

\begin{longtable}{|l|l|}
\hline
\multicolumn{1}{|c|}{\textbf{SRR}} & \multicolumn{1}{c|}{\textbf{TAXA}} \\ \hline
\endfirsthead
\multicolumn{2}{c}%
{{\bfseries Table \thetable\ continued from previous page}} \\
\hline
\multicolumn{1}{|c|}{\textbf{SRR}} & \multicolumn{1}{c|}{\textbf{TAXA}} \\ \hline
\endhead
SRR492065 & human gut metagenome \\ \hline
SRR492066 & human gut metagenome \\ \hline
SRR492182 & human gut metagenome \\ \hline
SRR492183 & human gut metagenome \\ \hline
SRR492184 & human gut metagenome \\ \hline
SRR492185 & human gut metagenome \\ \hline
SRR492186 & human gut metagenome \\ \hline
SRR492187 & human gut metagenome \\ \hline
SRR492188 & human gut metagenome \\ \hline
SRR492189 & human gut metagenome \\ \hline
SRR492190 & human gut metagenome \\ \hline
SRR492191 & human gut metagenome \\ \hline
SRR492193 & human gut metagenome \\ \hline
SRR492194 & human gut metagenome \\ \hline
SRR492195 & human gut metagenome \\ \hline
SRR492196 & human gut metagenome \\ \hline
SRR492197 & human gut metagenome \\ \hline
\end{longtable}

\end{document}